# Exactly Conservative Integrators

B. A. Shadwick, John C. Bowman, and P. J. Morrison

*Department of Physics and Institute for Fusion Studies*
*The University of Texas at Austin, Austin, TX 78712–1081*

August 2, 1995

**Abstract**

Traditional numerical discretizations of conservative systems generically yield an artificial secular drift of any nonlinear invariants. In this work we present an *explicit* nontraditional algorithm that exactly conserves these invariants. We illustrate the general method by applying it to the three-wave truncation of the Euler equations, the Lotka–Volterra predator–prey model, and the Kepler problem. This method is discussed in the context of symplectic (phase space conserving) integration methods as well as nonsymplectic conservative methods. We comment on the application of our method to general conservative systems.

# I. Introduction

For many years now symplectic integrators have been the subject of much productive study. (See Channell and Scovel[1] for an overview; see also the recent book by Sanz-Serna and Calvo[2].) There are variety of Hamiltonian systems for which symplectic methods have proven extremely useful, if not essential; but these methods do not constitute the last word on integration techniques. As Ge and Marsden show,[3] exact energy conservation is, in general, not possible with a symplectic method. Since the energy error is typically not secular but rather oscillatory, it is commonly believed that exact energy conservation is not as important a benefit as preserving the phase space structure.

Concerning the numerical preservation of more general constants of motion, less is known. Based on the work of Cooper,[4] Sanz-Serna[2,5] has shown that a restricted class of quadratic invariants will be conserved by certain symplectic Runge–Kutta schemes. For the Runge–Kutta methods studied by Cooper,[4] conservation of quadratic invariants necessarily requires that the method be implicit. One method of ensuring the preservation of any constant of motion is to use the constant to reduce the number of equations that must be solved. If the constants are in involution, then an entire degree of freedom (one coordinate and one momenta) can be removed from the dynamics for each such constant. This is seldom practical since the relationship between the constants of motion and a given dynamical variable may well be noninvertible (see the discussion in Gear[6]). The net result is that the reduced equations tend to be more complicated than the original system (hence the "force" terms are more expensive to compute); thus, in a system with a large number of degrees of freedom, little advantage is gained. Furthermore, if the constants of motion are not in involution, the system obtained by eliminating these invariants will be noncanonical,[7,8] resulting in even greater complexity.

It may be that the system of interest is most naturally described by variables that give rise to a noncanonical Hamiltonian structure. For noncanonical systems, Ge and Marsden[3] have provided a general construction for integrators that preserve both momentum maps and the structure of the Poisson manifold. Channell and Scovel[9] have shown how to implement these algorithms without the need of coordinatizing the configuration space group. This notwithstanding, they report that, with the exception of certain special (albeit important) forms of the Hamiltonian, such methods tend to be computationally expensive.

There is a further class of dynamical systems that is of interest, namely those systems that are not Hamiltonian (canonical or otherwise) but still possess constants of motion. Prime examples of such systems are transport equations, such as the Boltzmann equation. Further examples are truncations of the Fourier-transformed Euler fluid equations. The untruncated equations constitute an infinite-dimensional Hamiltonian field theory, however, when the number of Fourier modes is reduced to a finite set, the Hamiltonian structure is typically lost even though energy and enstrophy are still conserved. One might (justifiably) argue that such a truncation is not appropriate, but presently there is no practical alternative. (It is very much an open question as to the overall effect on the dynamics of such truncations.) Given that these systems are not Hamiltonian, symplectic methods are of no use, while the preservation of constants of motion is still of great interest.

A variety of methods for enforcing conservation of general invariants has been proposed. Baylis and Isaacson[10,11] have proposed a two stage algorithm where the approximate solution, obtained by standard methods in the first stage, is projected onto the constraint surface defined by the invariants in the second stage. Brasey and Hairer[12] have proposed a "half-explicit" method where the projection (*via* a Lagrange multiplier) and integration stages are merged together. LaBudde and Greenspan[13,14] have developed an algorithm for central force problems that conserves both energy and angular momentum. Gear[6,15] advocates an approach that amounts to an embedding of the original system into a higher dimensional space, yielding a set of differential-algebraic equations, the solution of which coincides with the solution of the original equations and preserves the invariants.

Our purpose in this paper is to present another approach to the development of exactly conservative algorithms. We begin with a simple model problem, that is of interest in both fluid mechanics and plasma physics, which possess two quadratic invariants. We develop integrators for this system that are *explicit* and exactly conserve both invariants. We further illustrate our method by applying it to the Lotka–Volterra predator–prey model and to the Kelper problem.

# II. A Model Problem

Our original interest in the issues of exact preservation of constants of motion arose in the study of two-dimensional inviscid fluid turbulence. As an illustration, consider the "three-wave" problem obtained by restricting the Fourier-transformed Euler equations to three modes[16–18]:

$$\frac{d\psi_K}{dt} = M_K \psi_P \psi_Q \equiv S_K(\psi), \tag{1a}$$

$$\frac{d\psi_P}{dt} = M_P \psi_Q \psi_K \equiv S_P(\psi), \tag{1b}$$

$$\frac{d\psi_Q}{dt} = M_Q \psi_K \psi_P \equiv S_Q(\psi), \tag{1c}$$

where $\psi = (\psi_K, \psi_P, \psi_Q)$, $K$, $P$, and $Q$ are the magnitudes of the Fourier wavenumbers of the three modes and the mode coupling coefficients $M_K$, $M_P$ and $M_Q$ satisfy

$$M_K + M_P + M_Q = 0, \tag{2}$$

and

$$K^2 M_K + P^2 M_P + Q^2 M_Q = 0. \tag{3}$$

This system possess two invariants: the total energy

$$E = \frac{1}{2}\left(\psi_K^2 + \psi_P^2 + \psi_Q^2\right) \tag{4}$$

and the total enstrophy

$$Z = \frac{1}{2}\left(K^2 \psi_K^2 + P^2 \psi_P^2 + Q^2 \psi_Q^2\right). \tag{5}$$

The constancy of these quantities follows directly from properties of $S_k$:

$$\sum_k \psi_k S_k = 0, \tag{6a}$$

$$\sum_k k^2 \psi_k S_k = 0, \tag{6b}$$

where $k$ ranges over the set $\{K, P, Q\}$. (These equations are isomorphic to Euler's equations for the rigid body, in which case the second invariant is the total angular momentum.)

When (1) is integrated numerically using standard methods neither $E$ nor $Z$ are exactly conserved. This behavior is made apparent by applying Euler's method with a time step $\tau$:

$$\psi_k(t+\tau) = \psi_k(t) + \tau S_k, \qquad k \in \{K, P, Q\}. \tag{7}$$

The energy at the new time is

$$\begin{aligned} E(t+\tau) &= \frac{1}{2} \sum_k \left[ \psi_k(t) + \tau S_k \right]^2 \\ &= \frac{1}{2} \sum_k \left[ \psi_k^2 + 2\tau S_k \psi_k + \tau^2 S_k^2 \right] \\ &= E(t) + \frac{1}{2} \tau^2 \sum_k S_k^2, \end{aligned} \tag{8}$$

where we have used (6a) in the last step. Thus the total energy is *always* increasing. A similar calculation for the enstrophy gives

$$Z(t+\tau) = Z(t) + \frac{1}{2} \tau^2 \sum_k k^2 S_k^2, \tag{9}$$

which is likewise always increasing. For extremely long runs these results imply that a very small time step is required to keep the accumulated error down to a given level — clearly an undesirable situation.

Many authors have noted that the lack of preservation of constants of motion potentially introduces significant unphysical effects and as such these errors are, in some sense, more important than those numerical errors that do not alter constants of motion. As de Frutos and Sanz-Serna[19] point out, one can think of the local error in a numerical integration as having two "components": one which leads to unphysical changes in the constants of motion and another which does not. When these local errors accumulate over many time steps, the former component is significantly more harmful than the latter in that errors which lead to changes in the constants of motion affect the *qualitative* nature of the solution, whereas other errors only affect the *quantitative* results. (A similar observation regarding the accumulation of error in area-preserving maps has been made by Greene.[20]) In essence we are saying that nonconserving integrators have the potential to make "structural" errors in the solution — an observation which agrees well with one's physical intuition. In the context of our model problem the implication is clear: keeping the time step small enough to maintain a reasonable level of energy and enstrophy conservation means that we are likely to be using more computational resources to obtain a given accuracy in the solution than would otherwise be necessary with a conservative integrator.

Although the three-wave problem is both integrable and Hamiltonian, our ultimate interest in this problem concerns the $n$-wave generalization of this system, which possesses both energy and enstrophy invariants but is not Hamiltonian; hence, we are lead to consider methods that do not rely on a particular geometrical structure. One might be tempted to enforce energy and enstrophy conservation by using these invariants to eliminate two modes from the dynamics. In this case the algebraic relations are simple enough to allow this, but there is a compelling physical argument against this approach. The modes are associated with different length scales; the choice of which modes to eliminate in favor of the invariants therefore has significant physical implications. Furthermore, the numerical error responsible for the lack of energy conservation could be imagined to contribute to a nonphysical energy and enstrophy transport. In any event, since each mode makes a *positive* contribution to both the energy and enstrophy, this scheme would eventually fail when the artificial growth in the invariants surpasses the initial contributions of the eliminated modes. Doubtless this scheme would exhibit nonphysical behavior long before this stage of failure was reached.

## III. Conservative Integrators for the Model Problem

In light of the above discussion, an algorithm that exactly conserves energy and enstrophy is clearly desirable. As we have noted in Section I, a variety of implicit methods are known that preserve quadratic invariants. While implicit methods have noteworthy stability properties, they tend to be less computationally efficient than explicit methods since they typically require multiple evaluations of the "force" terms. Therefore, we turn our attention to the development of explicit conservative methods for our model problem.

An elegant approach to this problem is found by borrowing from the ideas of backward error analysis.[2] The essential idea is to construct a new system of equations that, under the conventional (nonconservative) integrator, yields a conservative numerical approximation to the original equations. To this end, consider the alternative problem described by three equations of the form

$$\frac{d\psi_k}{dt} = S_k(\psi) + f_k. \tag{10}$$

Our objective is to find an $f_k$ that guarantees exact energy and enstrophy conservation and that vanishes in the limit of small step size. The form of $f_k$ will depend on the integration algorithm. We begin by deriving $f_k$ for Euler's method. We then construct a second-order predictor–corrector scheme.

## A. Euler's Method

As a "proof of principle" test we develop a conservative version of Euler's method. While not particularly useful in practice, Euler's method has the advantage that the algebra associated with constructing the conservative method is quite straightforward.

Application of Euler's method to the modified system yields

$$\psi_k(t+\tau) = \psi_k(t) + \tau(S_k + f_k). \tag{11}$$

The energy at the new time,

$$\begin{aligned} E(t+\tau) &= \frac{1}{2}\sum_k \left[\psi_k(t) + \tau\left(S_k + f_k\right)\right]^2 \\ &= E(t) + \frac{1}{2}\sum_k \left[2\tau f_k \psi_k + \tau^2(S_k + f_k)^2\right], \end{aligned} \tag{12}$$

will be conserved provided

$$\sum_k \left[2\tau f_k \psi_k + \tau^2(S_k + f_k)^2\right] = 0. \tag{13}$$

There is considerable freedom in satisfying (13). To guarantee that our approximate solution satisfies the original differential equation, it is necessary that $f_k$ vanish as $\tau \longrightarrow 0$. That is, for small time steps, we must recover the original integration algorithm. One would prefer that $f_k$ not introduce additional couplings into the differential equations. In light of these observations, let us try to satisfy (13) with the more restrictive condition that each term in the sum must independently vanish:

$$2 f_k \psi_k + \tau\left(S_k + f_k\right)^2 = 0. \tag{14}$$

There is an additional motivation for splitting (13) into three equations, namely that for $f_k$ satisfying (14), the enstrophy will also be conserved. This equation is easily solved, yielding

$$\tau f_k = -(\psi_k + \tau S_k) + \sigma_k \sqrt{\psi_k^2 + 2\tau S_k \psi_k}. \tag{15}$$

where $\sigma_k \equiv \sigma_k(t,\tau)$ is so far an unknown sign. Evaluation of (15) at $\tau = 0$ implies that $\sigma_k(t,0) = \text{sgn}(\psi_k(t))$. Upon substituting (15) into the Euler integrator, (11), we obtain the following time stepping rule:

$$\psi_k(t+\tau) = \sigma_k \sqrt{\psi_k^2 + 2\tau S_k \psi_k}. \tag{16}$$

It is now clear that $\sigma_k(t,\tau)$ must in fact be the sign of $\psi_k(t+\tau)$.

If $\psi_k(t) \neq 0$, then for sufficiently small $\tau$ the sign can be expressed explicitly as $\sigma_k = \text{sgn}(\psi_k(t))$. In this limit, $f_k$ then vanishes, or equivalently, (16) reduces to Euler's method:

$$\psi_k(t+\tau) = \text{sgn}(\psi_k(t)) \sqrt{\psi_k^2 + 2\tau S_k \psi_k}$$
$$\approx \psi_k + \tau S_k. \tag{17}$$

In this case the new algorithm predicts values of $\psi_k(t+\tau)$ that are quite close to those given by Euler's method — this is exactly what one would expect. The energy and enstrophy errors arising from (7) are the result of small (but nontrivial) errors in $\psi_k(t+\tau)$ that can be corrected by making a slight modification to the algorithm.

However, if $\psi_k(t) = 0$, it is seen from (15) that $f_k$ has the nonzero limit $-S_k$ as $\tau \longrightarrow 0$. Consequently, (16) has a spurious fixed point at $\psi_k(t+\tau) = 0$. Fortunately, a remedy for this problem has been developed. Equation (16) may be used up to and including the time step where $\psi(t+\tau) = 0$. Likewise, a scheme that steps *backwards* in time from $t + 2\tau$ may be used to integrate all the way back to $t + \tau$, where the two solutions must match. In this manner, we have developed an algorithm for the case where $\psi_k(t)$ changes sign.

Another potential problem with (16) is that the argument of the radical can become negative. In this case, we can rewrite the radical term as $\sqrt{\psi_k \chi_k}$, where $\chi_k = \psi_k + 2\tau S_k$ is just the Euler approximation for a step size of $2\tau$. The condition $\psi_k \chi_k < 0$ tells us that Euler's method is predicting a sign change of $\psi_k$ between $t$ and $t + 2\tau$; hence, we are in the vicinity of $\psi_k = 0$. To deal with this case, one adjusts the time step so that at the new time $\psi_k = 0$ and then proceeds with the implicit algorithm described above.

We give the name "Conservative Euler" (C–Euler) to this combined algorithm. In Figure 1 we compare the numerical solutions of the three-wave problem obtained using the conventional Euler method with those obtained using C–Euler and with the exact solution. For these calculations $K = \sqrt{3}$, $P = 3$, $Q = \sqrt{6}$, $M_K = 1$, $M_P = 1$ and $M_Q = -2$. The effect of the energy growth can be seen on the amplitudes computed by the Euler method.

With the exception of the points where $\psi_k$ changes sign, C–Euler is an explicit algorithm. We will soon see that the gymnastics associated with the fixed point of the integrator just described are a consequence of the low order of the Euler method and that a fully explicit conservative integrator is possible.

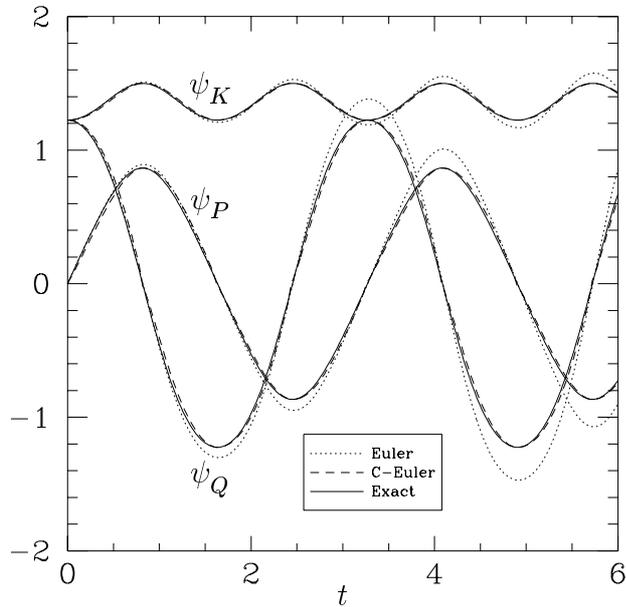

Figure 1: Solutions of the three-wave problem for the initial conditions $\psi_K = \sqrt{1.5}$, $\psi_P = 0.0$, and $\psi_Q = \sqrt{1.5}$ computed using the conventional Euler (Euler), the conservative Euler (C–Euler) and the exact solution (Exact). A fixed time step of 0.02 was used. The unphysical energy growth in the conventional Euler algorithm leads to large errors in the amplitudes.

## B.  Predictor–Corrector

In practice, one would prefer to use a scheme that both is of higher order than Euler's method and has better stability properties. We now turn to a simple second-order predictor–corrector scheme, which we apply to our model problem (1):

$$\widetilde{\psi}_k = \psi_k + \tau\, S_k \tag{18a}$$

$$\psi_k(t+\tau) = \psi_k + \frac{\tau}{2}\left(S_k + \widetilde{S}_k\right), \tag{18b}$$

where $\widetilde{S}_k = S_k(\widetilde{\psi})$. As we will show, using a second-order method overcomes the fixed-point problem that we encountered with Euler's method.

The energy now evolves according to

$$\begin{aligned}
E(t+\tau) &= \frac{1}{2}\sum_k \left[\psi_k(t)^2 + \tau\,\psi_k\left(S_k + \widetilde{S}_k\right) + \frac{\tau^2}{4}\left(S_k + \widetilde{S}_k\right)^2\right] \\
&= E(t) + \frac{1}{2}\sum_k \left[\tau(\psi_k\,S_k + \widetilde{\psi}_k\,\widetilde{S}_k) - \tau^2\, S_k\,\widetilde{S}_k + \frac{\tau^2}{4}(S_k + \widetilde{S}_k)^2\right] \\
&= E(t) + \frac{\tau^2}{8}\sum_k \left(S_k - \widetilde{S}_k\right)^2,
\end{aligned} \tag{19}$$

where we have used the definition of $\widetilde{\psi}_k$ and the properties of $S_k$ in the final step. A similar calculation gives

$$Z(t+\tau) = Z(t) + \frac{\tau^2}{8}\sum_k k^2 \left(S_k - \widetilde{S}_k\right)^2. \tag{20}$$

Again we see that the numerical method yields an ever increasing energy and enstrophy.[†]

To obtain a conservative version of this algorithm, we proceed as above by applying the predictor–corrector method to the modified equation of motion, (10), giving

$$\widetilde{\psi}_k = \psi_k + \tau\left(S_k + f_k\right), \tag{21a}$$

$$\psi_k(t+\tau) = \psi_k + \frac{\tau}{2}\left(S_k + f_k + \widetilde{S}_k + \widetilde{f}_k\right). \tag{21b}$$

---

[†] One might be tempted to conclude that any conventional method will yield a positive-definite energy growth. While nonconservation is generic, the sign of the energy error is typically indefinite. For example, a second-order Runge–Kutta method gives oscillatory errors in energy and enstrophy, although on average both the energy and enstrophy grow.

As we commented above, the conservative algorithm makes only small corrections to the values of $\psi_k(t+\tau)$. This immediately brings to mind the underlying philosophy of the predictor–corrector algorithms; in fact, one might suspect that energy and enstrophy conservation can be achieved by modifying *only* the corrector part of the integrator. Since the predictor is merely an intermediate approximation, there is surely no need for it to be conservative. Thus we can replace (21) with the simpler prescription

$$\widetilde{\psi}_k = \psi_k + \tau S_k, \tag{22a}$$

$$\psi_k(t+\tau) = \psi_k + \frac{\tau}{2}\left(S_k + \widetilde{S}_k + g_k\right). \tag{22b}$$

As before, we determine $g_k$ by demanding conservation of energy and enstrophy. The energy at $t+\tau$ is given by

$$\begin{aligned} E(t+\tau) &= \frac{1}{2}\sum_k \left[\psi_k(t)^2 + \tau\psi_k\left(S_k + \widetilde{S}_k + g_k\right) + \frac{\tau^2}{4}\left(S_k + \widetilde{S}_k + g_k\right)^2\right] \\ &= E(t) + \frac{\tau}{2}\sum_k \left[g_k\psi_k - \tau S_k\widetilde{S}_k + \frac{\tau}{4}(S_k + \widetilde{S}_k + g_k)^2\right], \end{aligned} \tag{23}$$

where the last step follows from the definition of the predictor and the properties of $S_k$. We see that energy will be conserved provided that

$$\sum_k \left[g_k\psi_k - \tau S_k\widetilde{S}_k + \frac{\tau}{4}(S_k + \widetilde{S}_k + g_k)^2\right] = 0. \tag{24}$$

Similarly, enstrophy will be conserved if

$$\sum_k k^2\left[g_k\psi_k - \tau S_k\widetilde{S}_k + \frac{\tau}{4}(S_k + \widetilde{S}_k + g_k)^2\right] = 0. \tag{25}$$

We can satisfy these conditions simultaneously if we can solve

$$g_k\psi_k - \tau S_k\widetilde{S}_k + \frac{\tau}{4}(S_k + \widetilde{S}_k + g_k)^2 = 0 \tag{26}$$

for $g_k$. Some straightforward algebra gives

$$\frac{\tau}{2}g_k = -\left[\psi_k + \frac{\tau}{2}\left(S_k + \widetilde{S}_k\right)\right] + \sigma_k\sqrt{\psi_k^2 + \tau\left(\psi_k S_k + \widetilde{\psi}_k \widetilde{S}_k\right)}, \tag{27}$$

where we choose $\sigma_k = \pm 1$ such that as $\tau \longrightarrow 0$, $g_k$ vanishes. We consider the limit of small $\tau$ in two cases. If $\psi_k$ is nonzero, then for small enough $\tau$, both $\psi_k$ and $\widetilde{\psi}_k$ have the same sign and we can expand the radical to give

$$\frac{\tau}{2} g_k = -\psi_k - \frac{\tau}{2}\left(S_k + \widetilde{S}_k\right) + \sigma_k \operatorname{sgn}(\psi_k)\left[\psi_k + \frac{\tau}{2}\left(S_k + \widetilde{S}_k\right)\right] + O(\tau^2), \qquad (28)$$

leading us to choose $\sigma_k = \operatorname{sgn}(\psi_k)$. Otherwise, if $\psi_k = 0$, then $\widetilde{\psi}_k = \tau S_k$ and $\widetilde{S}_k = S_k + O(\tau)$, so that

$$\begin{aligned}\frac{\tau}{2} g_k &= -\tau S_k + \sigma_k \sqrt{\tau^2 S_k^2} + O(\tau^2) \\ &= -\tau S_k + \tau \sigma_k \operatorname{sgn}(S_k) S_k + O(\tau^2). \end{aligned} \qquad (29)$$

In this case we take $\sigma_k = \operatorname{sgn}(S_k) = \operatorname{sgn}(\widetilde{\psi}_k)$. In the previous case, we noted, for small $\tau$, that $\psi_k$ and $\widetilde{\psi}_k$ have the same sign. Therefore, the choice $\sigma_k = \operatorname{sgn}(\widetilde{\psi}_k)$ will always provide the correct limiting behavior.

Using the expression (29) for $g_k$ in our modified predictor–corrector algorithm, (22), we obtain the following conservative integrator:

$$\widetilde{\psi}_k = \psi_k + \tau S_k, \qquad (30a)$$

$$\psi_k(t+\tau) = \widetilde{\sigma}_k \sqrt{\psi_k^2 + \tau\left(\psi_k S_k + \widetilde{\psi}_k \widetilde{S}_k\right)}, \qquad (30b)$$

where $\widetilde{\sigma}_k = \operatorname{sgn}(\widetilde{\psi}_k)$. Unlike the C–Euler algorithm, this algorithm, which we call "conservative predictor–corrector," (C–PC), does not suffer from fixed points. It is still possible that the argument of the radical can become negative; however, this merely indicates that the step size is too large.

We now compare the numerical solutions of our model problem obtained with the conventional predictor–corrector method with those obtained from C–PC, (30). Our results are summarized in Figures 2–5. In Figure 2 we show $\psi_k(t)$ computed with both methods as well as the exact solution. The errors in the two approximate solutions are displayed in Figure 3. The conservative predictor–corrector is seen to yield a more accurate solution. In Figure 4 we plot $\Delta E = E(t) - E(0)$ and $\Delta Z = Z(t) - Z(0)$ for both methods.

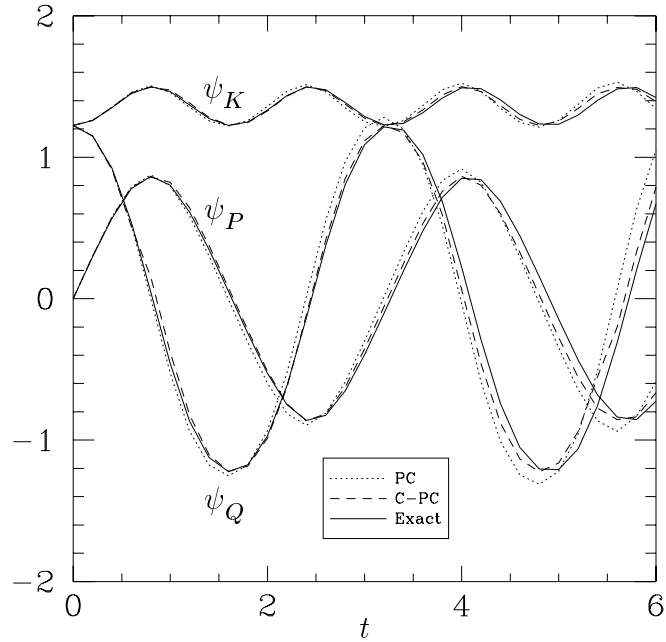

Figure 2: Solutions of the three-wave problem for the initial conditions $\psi_K = \sqrt{1.5}$, $\psi_P = 0.0$, and $\psi_Q = \sqrt{1.5}$ computed using the predictor–corrector (PC), the conservative predictor–corrector (C–PC), and the exact solution (Exact). A fixed time step of 0.2 was used.

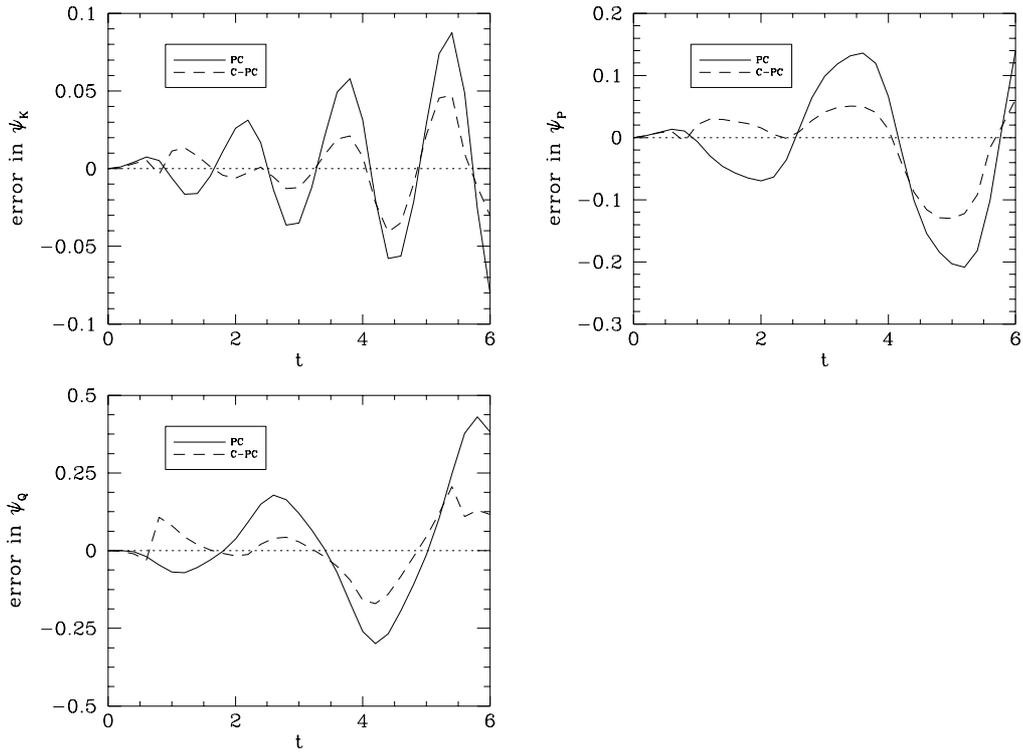

Figure 3: Differences between the computed and exact solutions in Figure 2.

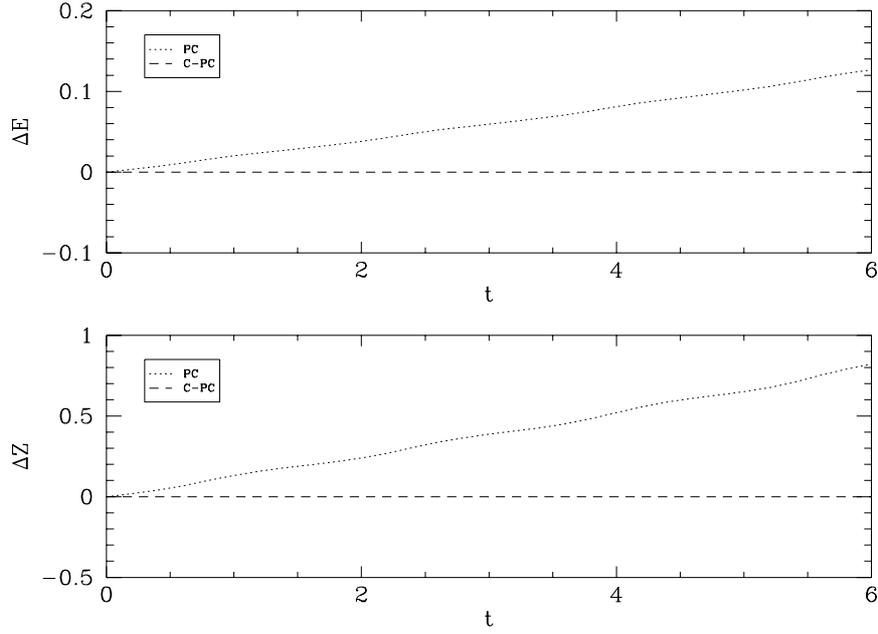

Figure 4: Change in energy and enstrophy for the conventional predictor–corrector method (PC) and the conservative predictor–corrector (C–PC).

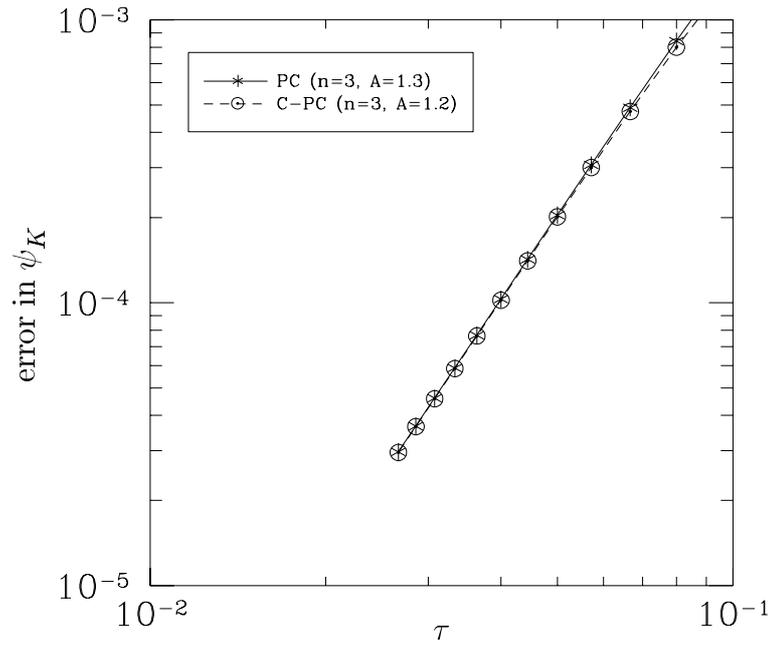

Figure 5: Single-step error in mode $K$ for the initial conditions $\psi_K = \sqrt{1.5}$, $\psi_P = 1.0$, and $\psi_Q = \sqrt{1.5}$ for the conventional and conservative predictor–corrector methods. The results of fitting the error to a power law $A\tau^n$ is shown, indicating that conservative algorithm is of second order, as expected.

In the limit of small step size, C–PC reduces to the conventional predictor–corrector. To illustrate this property we fit the error for a single step of mode $K$ to a power law:

$$\Delta \psi_k = A \tau^n. \tag{31}$$

The results of this fit, shown for both methods in Figure 5, are consistent with our expectation that both the conventional and conservative predictor–corrector methods are second order.

In constructing our conservative algorithms, we have essentially altered the manner in which truncation error enters the solution. Where this error has gone is an important question. It is unreasonable, of course, to expect that the truncation error has vanished. In fact, all of the truncation error is now lumped into the only place left — the phase of the solutions. Since our ultimate application is to fluid turbulence, the nature of this phase error could be of great importance. There are two general possibilities: either this error manifests itself as a global phase shift, with all three waves exhibiting the same phase error with respect to the exact solution, or each wave receives a different phase error, so that relative phase shifts begin to develop. Of the two possibilities, the first is of little consequence in a turbulence simulation, whereas the second could, arguably, be as bad (from a structural point of view) as the energy growth that we have sought to eliminate.

Since our model problem is integrable, we can easily distinguish between these cases. In a plot where each of the axes is one of the dynamical variables, an integrable system yields a simple closed curve. For our case, such a plot is shown in Figure 6. The solid line is the orbit computed with the conservative integrator while the dots represent the solution obtained from the conventional predictor–corrector. Since the conservative solution yields a closed curve, we may conclude that the additional phase error introduced is global and thus the relative phases of the waves are not affected by our method. This supports the general observation made by de Frutos and Sanz-Serna regarding the nature of local truncation error in systems with invariants.[19]

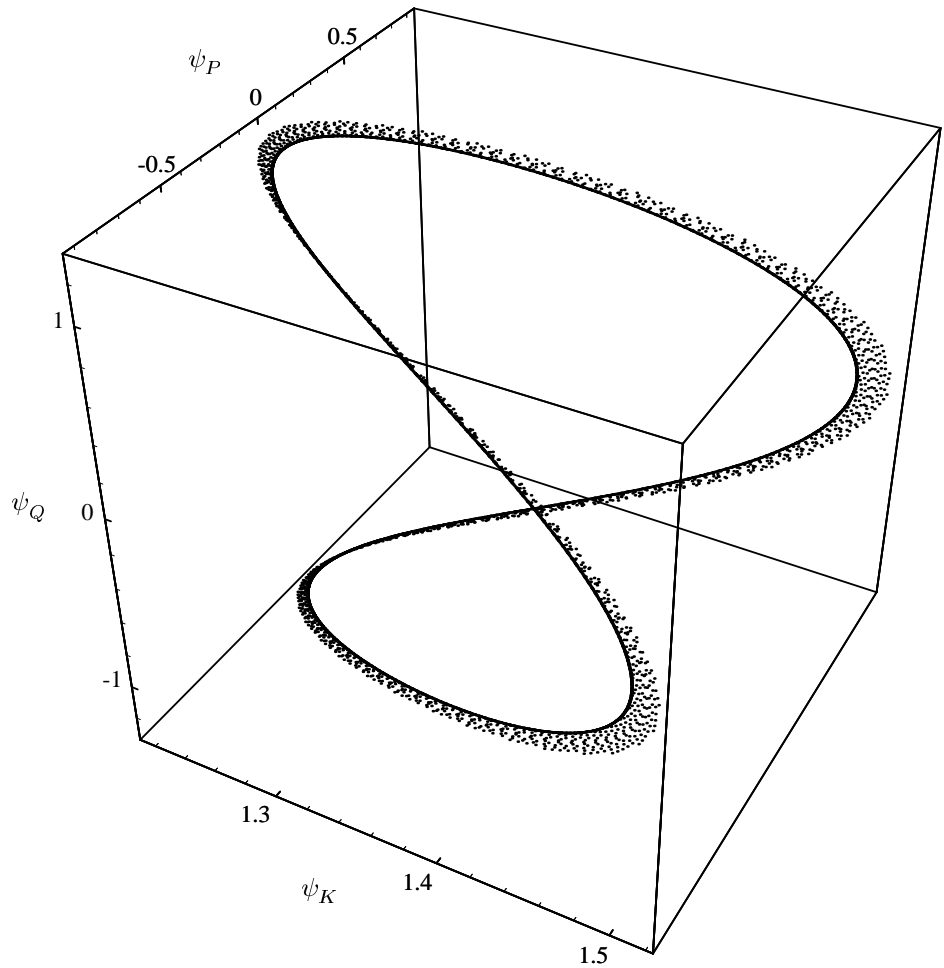

Figure 6: Integration of the three-wave problem using a conventional second-order predictor–corrector (solid line) and the conservative predictor–corrector (dots). The effect of the 4% energy gain by the conventional method is clearly visible.

## C. Generalizations

The C–PC algorithm has two important straightforward generalizations: to $n$-waves and to complex $\psi_k$. The $n$-wave generalization is immediate — nowhere in our derivations of the conservative algorithms have we made use of the number of modes. Both C–Euler and C–PC can be applied to a system with an arbitrary number of modes, where the energy and enstrophy expressions are the appropriate generalizations of (4) and (5) respectively.

The generalization to complex amplitudes proceeds as follows. Consider a system with $n$ complex-valued amplitudes $\psi_k$. We split these amplitudes into real and imaginary parts $\psi_k^r$ and $\psi_k^i$, respectively, which evolve according to

$$\frac{d\psi_k^r}{dt} = S_k^r(\psi), \tag{32a}$$

$$\frac{d\psi_k^i}{dt} = S_k^i(\psi), \tag{32b}$$

where $S_k^r$ and $S_k^i$ are the real and imaginary parts of the source function $S_k$. For this system the energy and enstrophy are given by

$$E = \frac{1}{2}\sum_k |\psi_k|^2 \tag{33}$$

and

$$Z = \frac{1}{2}\sum_k k^2 |\psi_k|^2, \tag{34}$$

where $k$ ranges over the wavenumbers of the $n$ modes. The properties of the source terms that guarantee conservation of energy and enstrophy are

$$\sum_k \psi_k^r S_k^r + \psi_k^i S_k^i = 0, \tag{35a}$$

$$\sum_k k^2 \left(\psi_k^r S_k^r + \psi_k^i S_k^i\right) = 0; \tag{35b}$$

hence, we see that a system of $n$ complex modes is completely equivalent to a system of $2n$ real modes. Therefore, the complex version of our conservative algorithms follows by applying the real algorithm separately to each component of the complex amplitudes.

# D. Discussion

It is worth saying a few words about computational efficiency. There are two sources of computational overhead associated with the conservative algorithms compared to the conventional methods. Here we concentrate on C–PC since C–Euler is not appropriate for practical use. In terms of operations, C–PC requires two additional multiplications and a square-root evaluation over the standard predictor–corrector method. Importantly C–PC uses no additional storage. The cost of the extra operations will, in most cases, be negligible compared to the cost of one evaluation of $S_k$. The square root is a cause for some concern as it may involve a function call. In any event, on modern hardware the square-root operation is only a small numbers of times (typically five to seven) slower than multiplication. Furthermore, it is reasonable to expect that a conservative integrator will obtain a given global accuracy with a larger time step than the corresponding conventional integrator, thereby ameliorating the overhead problem. The second source of overhead is the occasional need to reduce the time step when the argument of the square root becomes negative. In practice we find that this happens approximately 10% of the time and, in light of the above discussion, we feel that this is not significant.

Since the C–PC method reduces to the usual predictor–corrector algorithm in the small time-step limit, we expect that C–PC will inherit some of the stability properties of this method. While this does not establish the stability properties for large time steps, we find in practice that the C–PC method is numerically stable.

In numerical studies of the Euler fluid equations, an artificial viscosity is often added to the dynamical equations to compensate partially for the spurious growth of the energy and enstrophy introduced by the numerical scheme. The viscosity is usually taken to vary as a power of wavenumber. However, only one of the two invariants can be exactly conserved by such a procedure and even this requires that the prescribed viscosity coefficient be time dependent. Moreover, this remedy can be shown to contaminate the modal evolution. In contrast, the conservative algorithms developed in this work faithfully reproduce the modal dynamics.

In addition these methods can be applied to dissipative systems where the change in energy has a specific physical origin. The same numerical errors that previously led to nonconservation of energy will now contribute to the net energy change, thus having the effect of altering the underlying physics. For example, in a viscous fluid simulation the amount of energy leaving a mode is determined by the viscosity. It is a straightforward matter to use our methods to guarantee that the modal energy evolution is precisely that given by the physics. It is an open question and a subject of further investigation by the authors as to whether errors of this sort have the same structural effect on the solution as in the strictly conservative case.

There is a simple interpretation of both the C–Euler and C–PC algorithms that sheds light both on their form and on the existence of the two branches, labeled by $\sigma_k$. As numerous authors have observed, most traditional numerical methods conserve the linear invariants of a system. Consequently, one might be led to consider the possibility of transforming $\psi_k$ to new variables, in terms of which the invariants are linear. For the three-wave problem, this can be accomplished by making the transformation $\phi_k = \psi_k^2$. Upon applying the Euler method in the $\phi_k$ space and transforming back by taking the square root, one immediately obtains (16). This indicates that our restriction of the general constraint (13) to the condition (14) merely ensures that the modal energies evolve in a manner consistent with the Euler discretization of the energy equations. Below we will give derivations of integrators for other systems based on this idea. The C–PC algorithm can be viewed in the same light, except that the predictor is taken to have the simpler, nonconserving form. This also explains the $\psi_k = 0$ fixed point in C–Euler: the modal energies have a second-order zero at $\psi_k = 0$, thus it is no wonder that a first-order method fails at that point.

## IV. Lotka–Volterra

As a further demonstration, consider the Lotka–Volterra predator–prey equations:

$$\frac{dx}{dt} = -\mu x(1-y), \tag{36a}$$

$$\frac{dy}{dt} = y(1-x). \tag{36b}$$

These equations are surprisingly hard to integrate numerically since they are very susceptible to round-off error. With the exception of Kahan's[21] nontraditional method (which Sanz-Serna[22] has shown to be symplectic) there are virtually no other methods that can integrate this system without eventually failing due to round-off error.

This is a noncanonical Hamiltonian system[22] with Hamiltonian

$$H = x - \log x + \mu y - \mu \log y \tag{37}$$

and Poisson bracket

$$\{f, g\} = xy \left( \frac{\partial f}{\partial x} \frac{\partial g}{\partial y} - \frac{\partial g}{\partial x} \frac{\partial f}{\partial y} \right). \tag{38}$$

Just as with the three-wave problem, conventional integrators such as Euler and predictor–corrector fail to conserve total energy $H$. It happens that the dynamics of this system are particularly sensitive to the value of the energy, which explains the difficulty that these methods encounter when integrating (36).

It is possible to derive a conservative algorithm for this systems using the methods of backwards error analysis outlined above. The transcendental nature of the functions in the energy greatly complicates the procedure and prevents an analytical solution of the relevant equations. In light of these problems, we take an alternative approach to deriving a conservative integrator. We proceed using the observation that standard methods such as predictor–corrector exactly preserve linear invariants of a system of differential equations. To exploit this behavior, we introduce new variables $\xi_1$ and $\xi_2$ defined by

$$\xi_1 = x - \log x, \tag{39a}$$

$$\xi_2 = \mu \left( y - \log y \right). \tag{39b}$$

This transformation was chosen so that $H$ is a linear function of $\xi_1$ and $\xi_2$. Using the original equations of motion, we obtain

$$\frac{d\xi_1}{dt} = \mu(x-1)(y-1), \tag{40a}$$

$$\frac{d\xi_2}{dt} = -\mu(x-1)(y-1). \tag{40b}$$

Applying the usual second-order predictor–corrector to these equations yields

$$\widetilde{\xi}_1 = \xi_1 + \tau \mu (x-1)(y-1), \tag{41a}$$

$$\widetilde{\xi}_2 = \xi_1 - \tau \mu (x-1)(y-1), \tag{41b}$$

$$\xi_1(t+\tau) = \xi_1 + \frac{\tau}{2} \mu \left[ (x-1)(y-1) + (\widetilde{x}-1)(\widetilde{y}-1) \right], \tag{41c}$$

$$\xi_2(t+\tau) = \xi_2 - \frac{\tau}{2} \mu \left[ (x-1)(y-1) + (\widetilde{x}-1)(\widetilde{y}-1) \right]. \tag{41d}$$

Strictly speaking, here $\widetilde{x}$ and $\widetilde{y}$ are to be computed from $\widetilde{\xi}_1$ and $\widetilde{\xi}_2$ by inverting (39a) and (39b) respectively. Following the philosophy of the previous section, we instead compute $\widetilde{x}$ and $\widetilde{y}$ from the original equations of motion to obtain the following conservative integrator:

$$\widetilde{x} = x - \tau \mu x (1 - y), \tag{42a}$$

$$\widetilde{y} = y + \tau y (1 - x), \tag{42b}$$

$$\xi_1(t + \tau) = \xi_1 + \frac{\tau}{2} \mu [(x - 1)(y - 1) + (\widetilde{x} - 1)(\widetilde{y} - 1)], \tag{42c}$$

$$\xi_2(t + \tau) = \xi_2 - \frac{\tau}{2} \mu [(x - 1)(y - 1) + (\widetilde{x} - 1)(\widetilde{y} - 1)]. \tag{42d}$$

Here $x(t+\tau)$ and $y(t+\tau)$ are determined from $\xi_1(t+\tau)$ and $\xi_2(t+\tau)$ by inverting (39). Since this inversion requires solving a transcendental equation, in practice it will have to be carried out iteratively. Although the expressions for $x(t+\tau)$ and $y(t+\tau)$ can not be written in closed form, (42) is still an explicit scheme.

Notice that $\xi_1$ and $\xi_2$ are not one-to-one functions of $x$ and $y$; $\xi_1$ has a minimum value of 1 at $x = 1$, while $\xi_2$ has a minimum of $\mu$ at $y = 1$. These minimum values play the same role as the point $\psi_k = 0$ in the three-wave problem. Fortunately, the remedy is similar also: if either $\xi_1$ or $\xi_2$ are pushed below their respective minima, this indicates that the time step is too large. Temporarily reducing the time step alleviates this problem.

To illustrate the effectiveness of our conservative algorithm, we integrate (36) taking $\mu = 1.5$ with an initial condition of $x(0) = 1.0$ and $y(0) = 0.4$. In Figure 7 we show a comparison between the standard predictor–corrector and C–PC. The C–PC orbit exactly conserves energy and forms a closed curve. The predictor–corrector orbit spirals outward; a consequence of its energy gain.

We provide this example to illustrate the generality of our method; however, since there is not an explicit expression for the inverse of the transformation (39) and a symplectic algorithm is known,[21,22] this method seems to be of little practical value due to the computational overhead of iteratively determining $x$ and $y$ from $\xi_1$ and $\xi_2$.

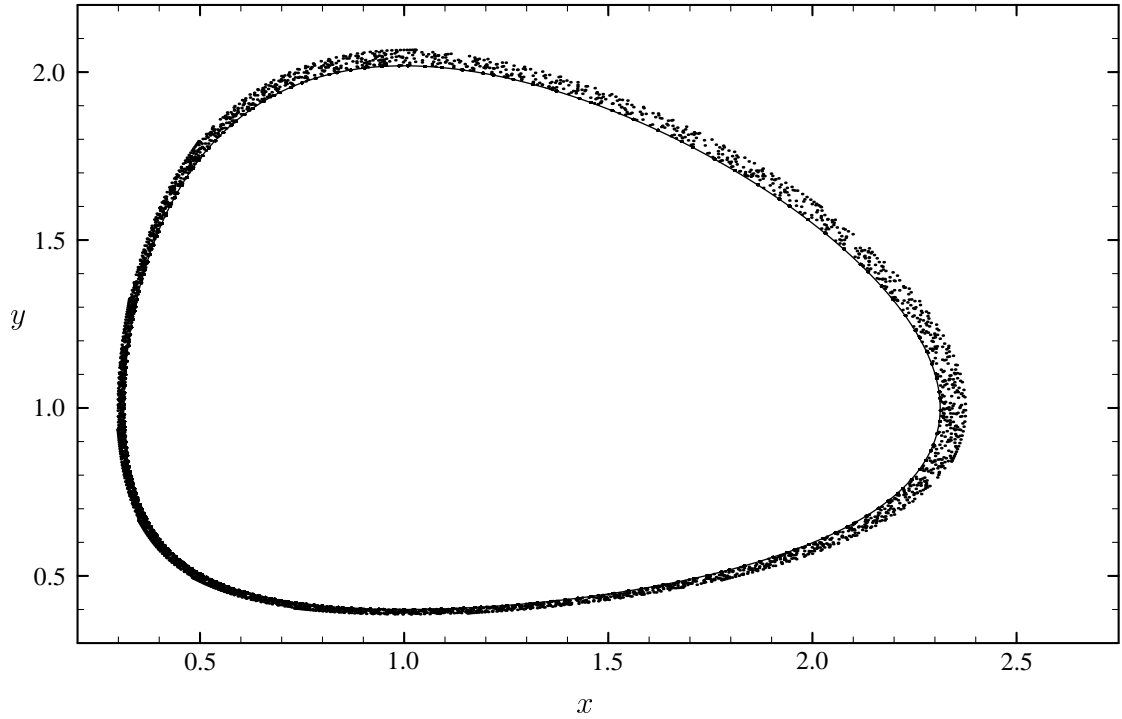

Figure 7: Integration of the Lotka–Volterra problem using a standard second-order predictor–corrector and the C–PC algorithm each with $8 \times 10^5$ time steps of size 0.02. A point is plotted every 200 time steps. The solid line represents the energy surface containing the initial condition. The points obtained from C–PC all lie on this curve. The dramatic effect of the 1.2% energy gain by the standard algorithm is clearly visible.

# V. Kepler Problem

As a final example we consider the problem of a single particle moving in a gravitational potential.[23,24] Let $\boldsymbol{r}$ be the position vector of the particle of mass $m$ and $\phi(r)$, where $r = |\boldsymbol{r}|$, be the gravitational potential. The equations of motion for this system are

$$\frac{d\boldsymbol{r}}{dt} = \boldsymbol{v}, \tag{43a}$$

$$\frac{d\boldsymbol{v}}{dt} = -\nabla \phi. \tag{43b}$$

This is a conservative system with the Hamiltonian

$$H = \frac{1}{2} m \boldsymbol{v}^2 + \phi(r). \tag{44}$$

As with all central force problems the total angular momentum, $\boldsymbol{L} = m \boldsymbol{r} \times \boldsymbol{v}$, is conserved, confining the motion to the plane perpendicular to $\boldsymbol{L}$. We exploit this feature by aligning our coordinate system with the $\hat{\boldsymbol{z}}$ direction parallel to $\boldsymbol{L}$ and introducing polar coordinates $(r, \theta)$ in the plane perpendicular to $\boldsymbol{L}$.

In these coordinates the equations of motion become

$$\frac{dr}{dt} = v_r, \tag{45a}$$

$$\frac{dv_r}{dt} = \frac{\ell^2}{m^2 r^3} - \frac{1}{m} \phi'(r), \tag{45b}$$

$$\frac{d\theta}{dt} = \frac{l}{mr^2}, \tag{45c}$$

where $\ell$ is the magnitude of the angular momentum and the Hamiltonian can be written as

$$H = \frac{1}{2} m v_r^2 + \frac{\ell^2}{2mr^2} + \phi(r). \tag{46}$$

Unlike all other central force problems, the Kepler problem has an additional constant of motion known as the Runge–Lenz vector,

$$\boldsymbol{A} = \boldsymbol{v} \times \boldsymbol{L} + \phi \boldsymbol{r}. \tag{47}$$

This invariant is somewhat unique in that it is not the consequence of a symmetry of the equations of motion but is a result of the potential being proportional to $1/r$. Conservation of the Runge–Lenz vector can be associated with the fact that the orientation of the bound orbits of this system is fixed. It turns out the these orbits are elliptical and oriented with the major axis in the direction of $\boldsymbol{A}$. We say "associated" here since in central force problems with any other force law, the orientation of the bound orbits precesses. Furthermore, the Runge–Lenz vector is in some sense redundant: it is *not* needed to integrate the equation of motion, as there are already enough constants of motion to render the problem integrable.

We adopt the initial conditions $r(0) = r_0$ and $\theta(0) = v_r(0) = 0$, so that the vector $\boldsymbol{A}$ is in the $x$-direction. Writing the potential as $\phi(r) = -K/r$, where $K$ is a constant, we see that the magnitude of $\boldsymbol{A}$ is given by

$$A = \frac{\ell^2}{mr_0} - K. \tag{48}$$

## A. A Conservative Integrator for the Kepler Problem

The Kepler problem is an interesting example to consider in a study of conservative integrators, not only as a preliminary to studying multi-body problems (which are of astronomical significance), but also because of the Runge-Lenz vector. While this vector is functionally dependent on the Hamiltonian and on the angular momentum, exact conservation of these invariants neither guarantees conservation of the Runge–Lenz vector nor prevents the computed orbits from exhibiting a spurious precession. Hence numerical conservation of the Runge–Lenz vector is as much a structural issue as is conservation of energy.

We now illustrate a conservative predictor–corrector (C–PC) for integrating (45) that exactly conserves $H$ and $A$. The predictor is conventional:

$$\widetilde{r} = r + \tau\, v_r, \tag{49a}$$

$$\widetilde{v}_r = v_r + \tau \frac{1}{mr^2}\left(\frac{\ell^2}{mr} - K\right), \tag{49b}$$

$$\widetilde{\theta} = \theta + \tau \frac{\ell}{mr^2}. \tag{49c}$$

To obtain the corrector equations, we transform $(r, v_r)$ to the new variables

$$\xi_1 = -\frac{K}{r}, \tag{50a}$$

$$\xi_2 = \frac{1}{2} m v_r^2 + \frac{1}{2} \frac{l^2}{mr^2}, \qquad (50b)$$

so that $H = \xi_1 + \xi_2$. Expressed in these new variables, the Hamiltonian is linear and will be conserved by conventional integrators. The corrector is given by

$$\xi_1(t+\tau) = \xi_1 + \Delta, \qquad (51a)$$
$$\xi_2(t+\tau) = \xi_2 - \Delta, \qquad (51b)$$

where

$$\Delta = \frac{\tau}{2}\left(\frac{Kv_r}{r^2} + \frac{K\widetilde{v}_r}{\widetilde{r}^2}\right). \qquad (52)$$

In terms of the original variables, (51) may be rewritten as

$$r(t+\tau) = \frac{-K}{-K/r + \Delta}, \qquad (53a)$$

$$v_r(t+\tau) = \mathrm{sgn}(\widetilde{v}_r)\sqrt{v_r^2 + \frac{L^2}{m^2}\left(\frac{1}{r^2} - \frac{1}{\widetilde{r}^2}\right) - 2\frac{\Delta}{m}}. \qquad (53b)$$

We still need an equation for $\theta$. Thus far, we have enforced the invariance of $H$, but not $\boldsymbol{A}$. Since only one integration variable remains to be determined, the conservation of $\boldsymbol{A}$ enforces the following constraint on $\theta$:

$$A\left(v_r \cos\theta - \frac{\ell}{mr}\sin\theta\right) = -K v_r, \qquad (54)$$

as is seen upon taking the $\boldsymbol{v}$-projection of (47). To avoid the complexities associated with multiply-branched solutions, the most efficient method for solving (54) appears to be Newton-Raphson iteration, using $\theta(t)$ for the initial estimate. Convergence is rapid; typically, only 3 or 4 iterations are required.

In Figures 8 and 9 we present our integration results for the conventional C–PC and PC algorithms, respectively, adopting the initial parameters $r = 1$, $v_r = \theta = 0$, $\ell = 1$, $K = 3/2$, and $m = 1$. To allow an even comparison, a slightly larger time step size was chosen for the C–PC run such that the amount of computer time needed to reach the final time was the same in both cases. The new algorithm dramatically outperforms the traditional integrator. The artificial precession of the trajectory exhibited by the predictor–corrector result does not occur in the C–PC solution, due to the explicit conservation of the Runge–Lenz vector.

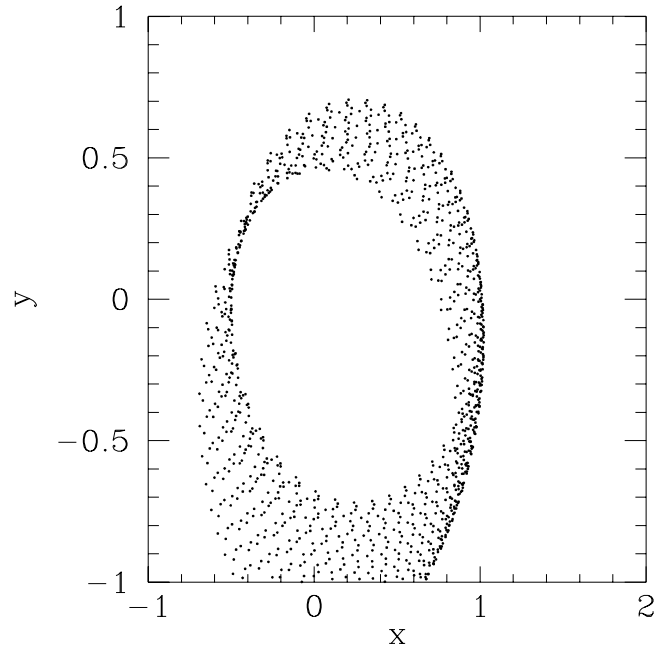

Figure 8: Solution of the Kepler problem computed using the conventional predictor–corrector. A total of 1313 fixed time steps of 0.08 were used.

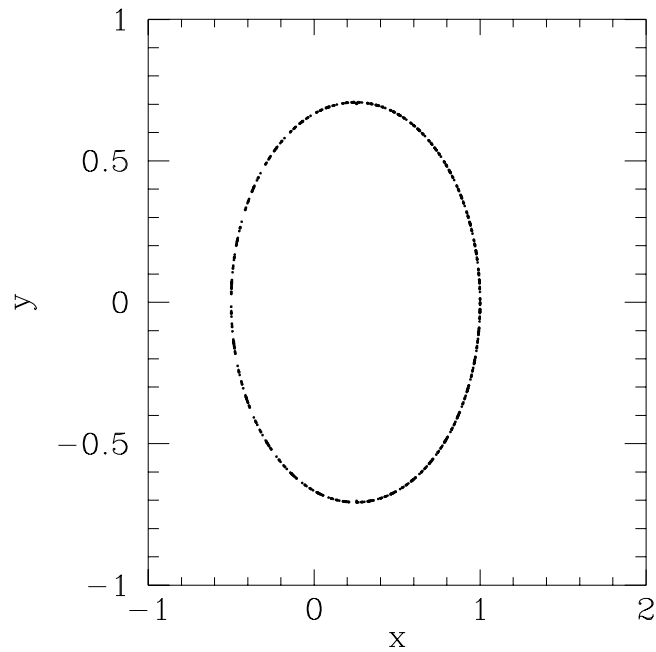

Figure 9: Solution of the Kepler problem computed using the conservative predictor–corrector. A total of 1000 fixed time steps of 0.105 were used.

## B. Discussion

We have demonstrated an explicit conservative integrator for the Kepler problem that captures all of its important structural features. LaBudde and Greenspan[13,14] have constructed integrators for this problem that conserve both energy and angular momentum but it is unclear whether their method exhibits orbital precession.

The extension of these ideas to multi-body problems is a complicated task. Although each of the components of the angular momentum are constants of motion, they are not in involution. In the simple Kepler problem we are able to avoid any difficulties associated with this behavior because the motion is confined to a plane perpendicular to the direction of the angular momentum. In the multi-body problem this is no longer the case, which significantly complicates matters.

## VI. Conclusions

We have demonstrated a method, based on the ideas of backwards error analysis, for deriving *explicit*, exactly conservative integration algorithms. This method consists of modifying the dynamical equations in such a way that when a particular conventional integration algorithm is applied to the modified equations, one obtains a solution consistent with the original equations that exactly conserves a system's invariants. This method will generally yield an explicit algorithm when an explicit conventional algorithm is chosen as the basis for the conservative scheme. We have seen that this method can be interpreted in terms of a transformation to a new set of variables in which the invariants in question are linear. This promises to be a general method for deriving conservative integrators.

In Section III, we saw that for a system with quadratic invariants, conservative integrators can be developed that are simple and computationally efficient. The case of quadratic invariants is of particular interest. The invariants in Lie–Poisson systems are typically quadratic

Casimirs. Furthermore, for Hamiltonian systems with Lie group symmetry, a Lie–Poisson system is the natural result of reduction; thus, our methods are applicable to integrating the dynamics on the Poisson manifold of such systems. For the integration of canonical Hamiltonian systems where the configuration space is a Lie group Simo, Lewis and co-workers[25–28] have developed a series of methods that are symplectic and conserve momentum. One could imagine a hybrid of these algorithms: a conservative integrator of the type discussed above for integrating the dynamics on the Poisson manifold coupled to the algorithms of Simo *et al.* for reconstruction the full phase space flow.

In addition to the desirable physical aspects of exact energy conservation there is some evidence[25] that such conservation leads to nonlinear numerical stability of the algorithm. Furthermore, any conservative integration method developed for general systems could certainly be applied to Hamiltonian systems, providing an interesting comparison with symplectic methods. For example, this might shed some light on the choice between preserving phase space structure and exact conservation of energy. In fact, one could envision using the local change in phase space volume as a diagnostic of the performance of a conservative integrator. These ideas will be the subject of a future paper.

# Acknowledgments

The authors would like to thank G. Tarkenton and R. Fitzpatrick for helpful conversations while developing these methods. This work was supported by the U.S. DoE under contract No. DE-FG05-80ET-53088.